# Presence of a Spatially Varying Electric Field at Lipid-Water Interface with Na/K ratio in Water

Biplab Bawali[1], Shubhadip Chowdhury[2], Smita Mukherjee[3], Angelo Giglia[4], Nicola Mahne[4], Stefano Nannarone[4], Mrinmay Mukhopadhyay[2], Jayashree Saha[1], and Alokmay Datta[1]

[1]Department of Physics, University of Calcutta, 92 Acharya Prafulla Chandra Ray Road, Kolkata 700 009, INDIA

[2]Surface Physics and Materials Science Division, Saha Institute of Nuclear Physics, 1/AF Bidhannagar, Kolkata 700 064, INDIA

[3]Department of Physics, Gobardanga Hindu College, Khantura, West Bengal 743 273, INDIA

[4]CNR-IOM – Istituto Officina dei Materiali, National Research Council of Italy, AREA Science Park, S.S. 14 Km 163.5, Basovizza, I-34012 Trieste, ITALY

## Abstract

The ion-lipid interface in Langmuir monolayers of Dipalmitoylphosphatidylcholine (DPPC), a major phospholipid component of the cell membrane, on pure water and 100 mM solutions of $Na^+$ and $K^+$ at different $[K^+]/[Na^+]$ ($\alpha$), atom/atom ratios, were studied initially by Surface Pressure ($\pi$) versus Specific Molecular Area (A) isotherms. The values of $\alpha$ were chosen as 0 (no $K^+$), 0.43 ($[K^+]:[Na^+]$ = 30:70) and 1.0 ($[K^+]:[Na^+]$ = 50:50) These monolayers were then transferred onto hydrophilized Si(001) substrates and studied through X-Ray Reflectivity (XRR) and Near Edge X-ray Absorption Fine Structure (NEXAFS) spectroscopy at the O K-edge. The two-dimensional rigidity of the monolayer obtained from isotherm studies, was found to increase with $Na^+$ ions with respect to the pristine monolayer but fall drastically and non-linearly below the pristine value with introduction of the $K^+$ ions, as $\alpha$ was increased. Analysis of the XRR profiles provided the thickness, average electron density (aed) and the interfacial roughness of the phosphatidylcholine head group and the two hydrocarbon tails of the monolayers on Si (001), from which the angle ($\phi$) between the head and the tails was determined. This was also seen to increase considerably from the pristine monolayer with incorporation of $Na^+$ ions, drop fast with $\alpha$ = 0.43, then increase a small amount at $\alpha$ = 1.0, showing a marked non-linearity. From NEXAFS, it was found that a linear increase in the cation ratio towards K led to a nonlinear variation in the P=O bond energy and a weakening of the P-O bond energy, the latter becoming more pronounced with K ions, consistent with Fajans rule. Also a split in the C=O $\pi$-bond peak was observed at $\alpha$ = 1.0, which maybe explained through two different polarizations due to a spatially varying electric field.

These results cannot be explained with the model of a uniform electric field due to the cations, which would fall linearly with increase in the $K^+$ proportion, and rather suggest a structured field due a spatial variation in charge density in an interfacial layer of high ion concentration assembled by the counterionic attraction of the phosphatidylcholine head groups. Our results have important implications for the cell membrane, where such mixtures at high concentrations constitute the norm.

1. Introduction

Ions in aqueous solutions are generally considered to be at very low densities where the solvation spheres of water molecules act as dipolar screens. The forces between any two ions are strongly modified by the arrangement of the water dipoles around the ions and are generally weak relative to the thermal forces. Thus the ions are effectively free and their motions are controlled dominantly by the temperature of the aqueous medium. However, even in this condition the charge density of the ions plays an important role in deciding the arrangement of the water molecules in the solvation sphere. For example, the water molecules surrounding a $Na^+$ ion, which has high charge density, are more ordered than those surrounding a $K^+$ ion with lower charge density [1–7]

As the concentration of the ions increases the Coulomb forces become progressively less screened and at very high concentrations the counterions form structures such as molecular dipoles and complex ions [8] and the apparent volume per ion deviates largely from the free ion volume [9]. If, in addition, only cations or anions are concentrated selectively then at high concentrations they are expected to form ordered lattices through Coulomb repulsion [10]. However, to our knowledge, such systems have been studied only for one type of ion, i.e. with the same charge density everywhere and hence with a uniform mean electric field. Very little, if at all, analysis has been carried out on a dense mixture of two types of cations with different charge densities, which may give rise to an electric field with significant local variations.

Surprisingly enough, such a situation exists in one of the most common and important instance in the life sciences, namely the $Na^+/K^+$ mixture at the external and internal interfaces of the cell membrane. The phosphatidylcholine headgroups of the different lipids in the bilayer membrane are dipoles with the negative charge sitting on the free O atom attached through a single bond to the P atom in the phosphatidyl moiety [11,12]. These attract the $Na^+$ and $K^+$ ions to form a dense layer at the interface and the consequent structured electric field controls the physico-chemical behaviour of the cell membrane and requires to be understood in details.

The present communication reports a step towards this understanding using the simplest bio-mime of a Langmuir monolayer of Dipalmitoylphosphatidylcholine (DPPC), a major constituent of cell membranes, on pure water and 10 mM solutions of $Na^+$, and mixtures of $Na^+$ and $K^+$ at different $[K^+]/[Na^+]$ ($\alpha$), atom/atom ratios. We have first extracted the two-dimensional rigidity of the monolayers on water from Surface Pressure ($\pi$) versus Specific Molecular Area (A) isotherms, then deposited the monolayers on hydrophilized Si(001) and extracted the various structural details from their X-ray Reflectivity (XRR) profiles, which then yielded the angle $\phi$ between the phosphatidylcholine headgroup and the hydrocarbon tails. We investigated the bonds in the monolayers on Si (001) through Near Edge X-ray Absorption Fine Structure (NEXAFS) spectroscopy.

While the results for the pure Na$^+$ solution are consistent with a uniform electric field at the headgroup-water or headgroup-Si (001) interfaces, this is not tenable for the mixtures of the cations. In particular, the negative charge at the O atom shows a smearing or delocalization over the other two single bonded O atoms in the phosphatidyl group. This suggests a variation of the electric field around these atoms.

## 2. Experimental Methods
### 2.1. Isotherm Studies on Water

Dipalmitoylphosphatidylcholine (DPPC), obtained from Avanti Polar Lipids with a purity exceeding 99%, was dissolved in chloroform, resulting in a stock solution with a concentration of 1mg/ml. High-purity NaCl (≥99%) and KCl (≥99%), obtained from Sigma-Aldrich, were dissolved in deionized Millipore water with a resistivity of 18.2MΩ-cm, creating sub-phase solutions with varying ratios of NaCl and KCl concentrations but an overall fixed cation concentration of 10 mM. Specifically, according to our definition of $\alpha =$ [K$^+$]/[Na$^+$], solutions with values of $\alpha = 0$, 0.43 and 1.0 were prepared and used. Experiments were carried out at ambient temperature and unadjusted pH of water having a nominal value of $\sim 5.5$.

75 μL of DPPC stock solution was spread onto the sub-phase surface in a Langmuir trough (NIMA, UK) and was equilibrated for about 15 mins to form each of the Langmuir monolayers. Subsequently, the monolayers were compressed at a velocity of 2 cm²/min. Surface pressure $\pi$ (in mNm$^{-1}$), defined as the difference between $\gamma_0$, the surface tension of pure water, and $\gamma$, the monolayer surface tension, was measured with the consequent change in the mean molecular area ($A$, in Å$^2$) to yield the $\pi$-A isotherms of each monolayer. For each monolayer, compression-decompression-recompression isotherms were recorded to check reproducibility and hysteresis, which would indicate the presence or absence of any permanent or chemical change. Again, the whole process was repeated thrice with fresh monolayers to check overall reproducibility. Isotherms were measured for the pristine monolayer, i.e. without any ions in the sub-phase, and for $\alpha = 0$ (no K$^+$), 0.43 and 1.0.

### 2.2. X-ray Reflectivity (XRR) Studies on Si (001)

The Langmuir-Blodgett technique was employed to transfer the DPPC monolayers onto hydrophilized silicon substrates. Prior to the deposition, the silicon substrates underwent thorough cleaning and were rendered hydrophilic by immersion in a solution containing hydrogen peroxide (30% from Merck), ammonium hydroxide (30% from Merck), and ultra-pure water (in a volume ratio of 1:1:2, respectively) at a temperature of approximately 100°C for approximately 30 minutes. All depositions were carried out at a fixed surface pressure of 35mN/m, under ambient conditions and at the unadjusted pH of 5.5 of the aqueous subphase. The film deposition followed the conventional Langmuir-Blodgett method for hydrophilic substrates, involving a deposition process characterized by only an up-stroke.

X-ray reflectivity data of the monolayers on Si (001) were obtained utilizing Cu K$\alpha$ radiation with a wavelength of $\lambda = 1.54$Å. This was conducted using a sealed tube source within a Rigaku Diffractometer to derive the X-ray reflectivity profiles of the films from which electron density profiles (EDPs), specifically the depth profiles of average electron density (aed) within the films, were extracted.

### 2.3. X-ray Absorption Near Edge Spectroscopy (XANES) Studies on Si(001)

XANES measurements have been performed on the monolayers on Si at BEAR beamline of Elettra Synchrotron, Trieste, Italy [13].

Measurements were carried out at O K absorption edge in the range 530-600 eV, with 0.01 eV steps. The incident light was horizontally elliptically polarized and the incidence angle of the light with respect to the sample surface plane was kept fixed at 90°. The spectra were collected in Total Electron (TEY) mode measuring the current from the sample, and Fluorescence Yield (TFY) mode, measuring the current from an absolute Si diode close to the sample. The incident photon flux $I_0$ was measured using a calibrated Si diode AXUV-300. Data and $I_0$ have been normalized by monitoring the photocurrent from the last mirror of the beamline. These measurements were also used to calibrate the spectra in energy and compared with previously characterized energy benchmarks.

### 3. Results and Discussions
#### 3.1. Monolayer Rigidity

In this section, our focus lies in understanding the effect of $K^+/Na^+$ ratio on the mechanical characteristics of the lipid monolayer.

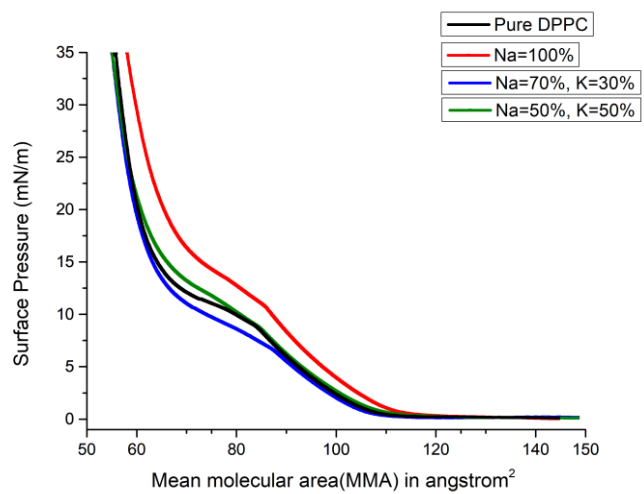

Figure 1. Surface Pressure ($\pi$, in mNm$^{-1}$) versus Mean Molecular Area ($A$, in Å$^2$) for Dipalmitoylphosphatidylcholine (DPPC) monolayers on pure water (DPPC), and on water containing 10 mM cations in [K$^+$]/[Na$^+$] ratios ($\alpha$) as 0 (100% Na$^+$), 0.43, 1.0. Data collected at ambient temperature and unadjusted pH of water ~ 5.5.

To address this, we undertake a series of isotherm experiments on DPPC lipid monolayers. These experiments are conducted both on pristine deionized water sub-phases and on sub-phases containing different ratios of the ions in solution. Figure 1 presents isotherms of the different monolayers as discussed above. They were reproducible and showed no hysteresis on decompression, clearly indicating the high degree of stability. Each of them show very similar sets of distinctive phases, beginning from the $L_d$ ('disordered liquid') phase at the lowest surface pressures and changing to the $L_o$ ('ordered liquid') phase through a quasi-first order transition beginning nearly at $\pi \sim$ 10 mNm$^{-1}$ and ending around $\pi \sim$ 15 mNm$^{-1}$. This is very much consistent with established behavior of stable and uncontaminated Langmuir monolayers of lipids [14–16].

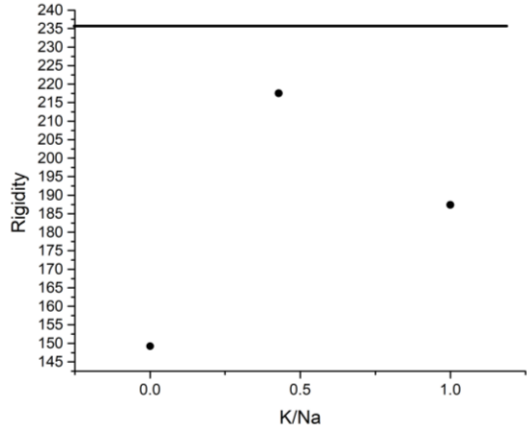

Figure 2. Variation of rigidity (β), defined as $-A\partial\pi/\partial A$ with K/Na ratio. Note that K/Na=0 indicates Na=100 and the solid line signify the corresponding value on pure water subphase

We have extracted the rigidity of the monolayers in the ordered liquid phase at a constant surface pressure of 30mN/m from the isotherms with rigidity defined as $\beta = -A\partial\pi/\partial A$ and have presented it versus the [K]/[Na] ratio in Figure 2. For lipid monolayers, rigidity quantifies the capacity of the monolayer to withstand deformation or alterations in form. The plot illustrates how the rigidity of the condensed phase evolves with the introduction of the ions.

We observe that the addition of $Na^+$ ions to the subphase causes a significant drop in the monolayer rigidity, reducing it by 86 $mNm^{-1}$ with respect to the rigidity on pure water. However, when $K^+$ ions are introduced to the subphase along with $Na^+$ ions in a 30:70 ratio, the rigidity remarkably increases by 68 $mNm^{-1}$ compared to the rigidity on water containing 100% $Na^+$. Interestingly, upon equalizing the concentrations of $Na^+$ and $K^+$ ions in the subphase, the monolayer rigidity decreases again relative to the rigidity observed with the 30:70 ionic ratio of $K^+$ and $Na^+$, respectively.

At a constant surface pressure, the different rigidity moduli indicate variations in the structural assembly of lipid molecules within the monolayer. As better packing leads to greater rigidity, and better packing can be achieved by straightening the molecules, it becomes evident that the addition of ions to pure water reduces rigidity, resulting in loosely packed monolayers. While $Na^+$ ions tend to fluidize the monolayer more [17], the incorporation of $K^+$ ions into the subphase probably causes the tail chains of the DPPC molecules to straighten, leading to improved packing and increased rigidity compared to the previous case with solely $Na^+$ ions.

### 3.2. Monolayer Angular Stress

On hydrophilized Si(001) substrates, we deposited the monolayers as Langmuir-Blodgett (LB) monolayer at various ionic ratios in ambient circumstances, and we explored their structural characteristics using X-Ray Reflectivity (XRR) analysis. In Figure3(a), we present the reflectivity profiles derived from our XRR experiments. These profiles were fit using the well-known Parratt method [18]. The results of these fits are depicted in Figure 3(b), showing the depth profiles of the average electron densities (aeds) of the films, namely the electron density profiles (EDPs). Through the subsequent analysis of electron density profiles, we determined the precise structural characteristics of the film. This analysis provided us with essential information regarding the electron density distribution and the individual thickness of each layer.

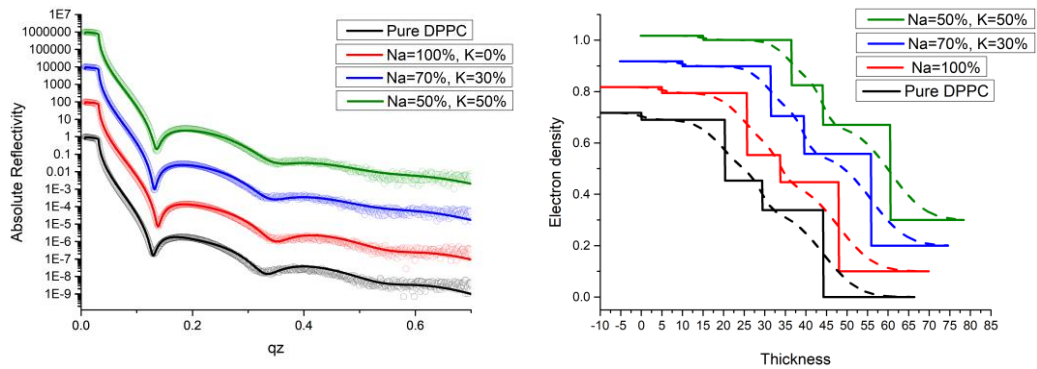

Figure 3. (a) X-Ray Reflectivity profiles of different lipid monolayer LB films deposited from monolayers on water containing different concentrations of Na and K ions, data in dots and Parratt fits in solid lines (b) the density profile obtained from reflectivity analysis of different lipid monolayer LB films, both the profiles are shifted towards up and right side by arbitrary amounts for visual clarity

The schematic diagram in Figure 4(b) offers an intuitive representation of the molecular organization within the film and explains the obtained key structural parameter values presented in Table 1. According to this electron density profile analysis, the DPPC lipid monolayer thickness over pure water is 23.9Å, which is identical to that obtained by Kienle et al. [19] using multiple-beam interferometry. It is clear that the presence of $Na^+$ and $K^+$ ions modify the monolayer thickness. The alterations in the thickness of both Layer-1 and Layer-2 are presumed to stem from the tilting of the head dipole and tail chain, respectively. These tilting movements are believed to be influenced by the Na/K ratio.

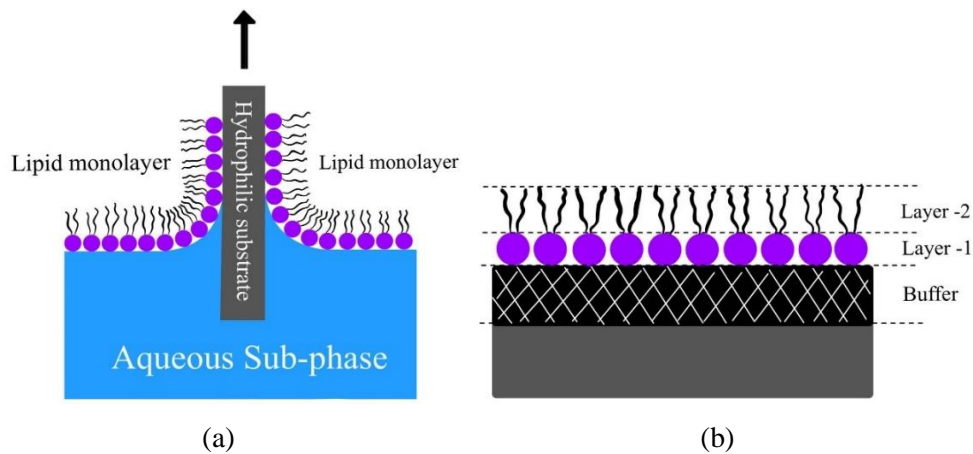

Figure 4. the suggested schematic diagram of the lipid monolayer film obtained from density profile analysis

To visualize the molecular structure, we introduced three angles - $\tau$, $\theta$, and $\phi$ - illustrated in Figure 5. Angle $\tau$ measures the tilt angle of the tail chain axis from the normal, while angle $\theta$ represents the tilt angle of the dipole within the head group of the lipid molecule from the normal. Tilt angles τ and θ have been calculated using the formula $\tau \text{ or } \theta = \cos^{-1}\left|\frac{observed\ thickness\ (l)}{maximum\ measured\ thickness\ (L)}\right|$ as shown in Figures 5(a) and (b). Angle ϕ (defined as $\phi = \pi - (\tau + \theta)$) characterizes the angle between the head dipole and the tail chain axis.

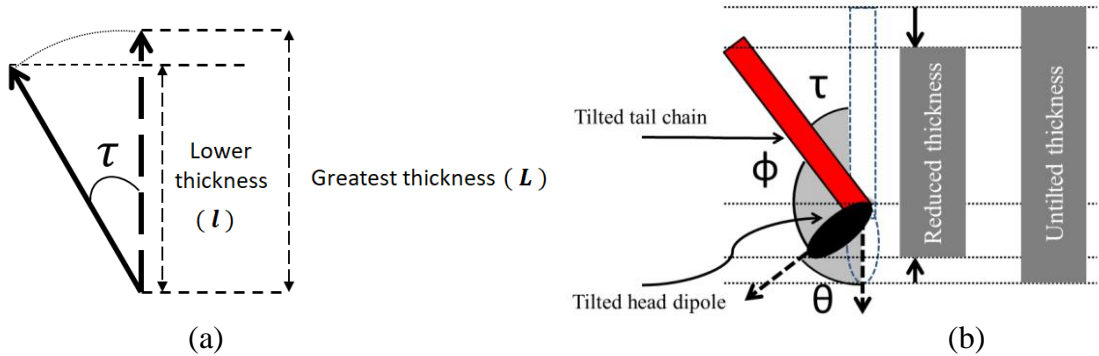

(a)                          (b)

Figure 5. Schematic diagram of molecule showcasing tilt angle of head dipole and tail chain axis along with the angle between head dipole and tail chain.

Based on the observed changes in thickness, we computed the tilt angles of the head dipole and tail chain within the lipid molecule. Then we determined the angle ϕ between the head dipole and the tail chain which is shown in Figure 6(a). Higher ϕ values imply straighter molecular configurations, resulting in a more efficient packing that yields lower mean molecular area. This leads to increased rigidity within the lipid bilayer. Conversely, lower ϕ values signify greater bending in the molecules, resulting in an expanded mean molecular area and contributing to the fluidity of the bilayer structure.

| samples | Layer-1 | | | Layer-2 | | |
|---|---|---|---|---|---|---|
| | Thickness | AED | Roughness | Thickness | AED | Roughness |
| Pristine | 9.1 | 0.45 | 2.0 | 14.8 | 0.34 | 6.8 |
| α = 0.0 | 8.1 | 0.45 | 2.0 | 14.2 | 0.35 | 6.8 |
| α = 0.43 | 8.1 | 0.50 | 2.0 | 16.3 | 0.36 | 6.8 |
| α = 1.0 | 7.6 | 0.52 | 2.0 | 16.4 | 0.37 | 6.8 |

Table1. Obtained value of thickness, density and roughness of each layers from XRR analysis

The plots in Figure 6 provide insights into the molecular-level changes induced by the presence of different ions in the sub-phase. On a pure water sub-phase, at a surface pressure of 35 mN/m, the angle between the head dipole and tail chain of the lipid molecules is observed to be around 145.4°. The corresponding monolayer on water has the maximum rigidity.

The introduction of solely $Na^+$ ions at 10 mM concentration to the sub-phase results in a significant reduction of this angle to 115.8° under the same surface pressure conditions. This decrease in the angle suggests a more disordered and fluid-like arrangement of the lipid molecules, corroborating the observed decrease in rigidity from our previous isotherm study.

When $K^+$ ions are introduced along with $Na^+$ ions, the angle between the head dipole and tail chain increases to 128.1°, indicating a greater degree of order and rigidity compared to the case with 100% $Na^+$. This finding agrees perfectly with our previous isotherm study, where the incorporation of $K^+$ ions led to an increase in the monolayer rigidity relative to the case with solely $Na^+$ ions present.

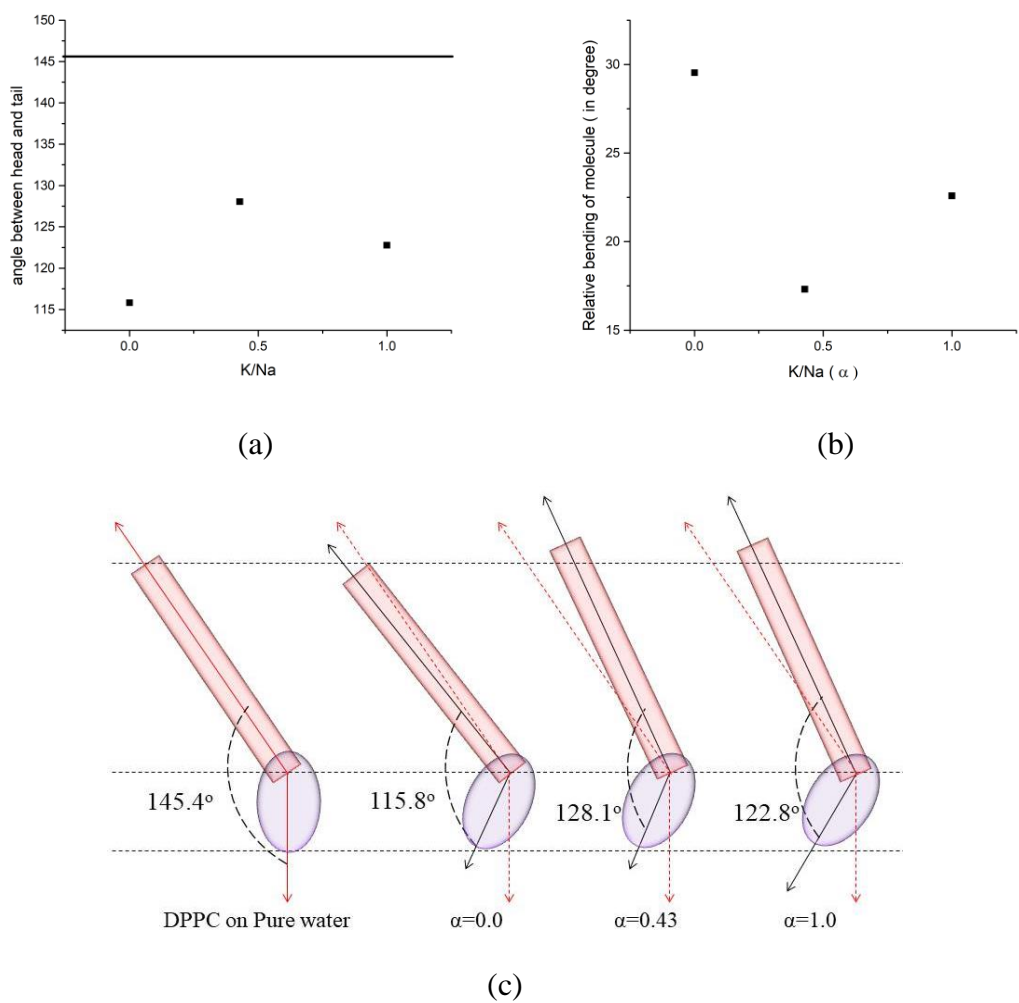

Figure 6. (a) plot of tilt angle of head dipole and tail chain axis; solid line is the respective tilt angle of lipid molecule on pure water subphase (b) relative bending of molecule with respect to Pure water(c) the schematic diagram of bending of DPPC molecule with ionic ratio

Furthermore, as the concentration of $K^+$ ions is increased to maintain an equal ionic ratio of 50:50 with $Na^+$, the angle slightly decreases to 122.8°, consistent with the expected behavior based on the rigidity curve presented in Figure 2. This observation highlights the direct correlation between the fluidizing effect of $Na^+$ ions and the rigidifying influence of $K^+$ ions on the lipid assembly with the angular stress within the lipid molecules, as involved with ϕ, induced by the ions. Other than this, the bending of the molecules as given by $\tau$ is shown in Figure 6(b) and the molecular geometries are schematically shown in Figure 6(c). It is clear that Na$^+$ ions tilt the molecule away from the normal obviously due the attraction of the head group to the ion, whereas introduction of the K$^+$ ions reduces this attraction.

The introduction of K ions reduces the charge density of the cation mixture due to its larger size. Since the K ion ratio is linearly increased, it is expected that the charge density and hence the electrostatic field due to these ions is also reduced linearly. However, from Figure 2 and Figure 6, it is apparent that the changes in rigidity and angular stress with ion ratio are strongly nonlinear. In particular, the situations at [K]/[Na] = 30:70, though mutually consistent, are anomalous and cannot be explained with an unstructured, dispersive distribution of the ions causing an uniform electrostatic field. This prompted us to investigate the bonding in the lipids, especially the head groups, as affected by the ions.

### 3.3. Bonding in the Monolayer

NEXAFS spectroscopic studies were carried out with four samples, viz. the DPPC sample (Figure 7(a)), and the DPPC samples with Na/K ions incorporated in the ratios Na:K=100:0 (Figure 7(b)), Na:K=70:30 (Figure 7(c)) and Na:K=50:50 (Figure 7(d)). O K-edge NEXAFS spectra of all films (solid circles) have been fitted (solid line) with voigt function. Fitting parameters are shown in Table 2. The low energy sharper feature of the spectra usually arise due to π* transitions whereas higher energy broader part denote σ* transitions [20], with deviations depending on nature of samples. Given the structure of DPPC, there are two types of bonding with oxygen, namely carbon-oxygen bonds and phosphorus-oxygen bonds. Peak assignments have been done, based on the presence of such bonds in samples.

Peak assignments are shown in Table 2, with all peaks numbered for clarity. Peak 1 at 535 eV, assigned to O 1s →π*$_{C=O}$ transition [21], was downshifted on addition of metal ions, and the shift was considerable for K ions, eventually splitting in two peaks at 534 eV (Peak 1) and 536 eV (Peak 2) for the Na:K=50:50 sample, probably indicative of two types of C=O bonds. Peak 4 at 549 eV for the DPPC sample, assigned to O 1s →σ*$_{C=O}$ transition [21], was downshifted to 547eV for Na:K= 100:0 sample and to 546.6 eV for the Na:K= 50:50 sample. This is in agreement with previous observation of downshift and splitting of the π* transition. There is also a notable decrease in intensity of peak 4 for DPPC samples with high Na concentration (Na:K=100:0 and Na:K=70:30). However, the peak intensity is found to be slightly more than pristine DPPC sample for Na:K=50:50. Peak 5 at 550 eV for the DPPC sample, assigned to P-O σ bonds [22,23], shows no change on incorporation of Na ion, but is downshifted on incorporation of K ions in the DPPC sample. However, there is an increase in intensity of peak 4 for higher Na concentration (Na:K=100:0 sample) but decreases gradually with increase in K ion incorporation. Peak 3 and peak 6 at 541.6 eV and 554.6 eV, respectively, for DPPC samples remain more or less unchanged for all samples. Peak 6 is assigned to P=O σ bonds [22,23], whereas peak 3 probably indicates either C-O transitions or Si-O bonds from the substrate. The sharp nature of this peak, unchanged in all samples hints to Si-O bonds, but given thickness of the sample and absence of any other peaks hints that peak 3 might be due to C-O transitions. The most important trend emerging from our results is that inclusion of Na/K ions strongly affect the C=O and P-O bonds in DPPC, as is evident from the downshift of peaks 4 and 5 at 549 eV and 550 eV, respectively, with incorporation and/or increase in Na/K concentration.

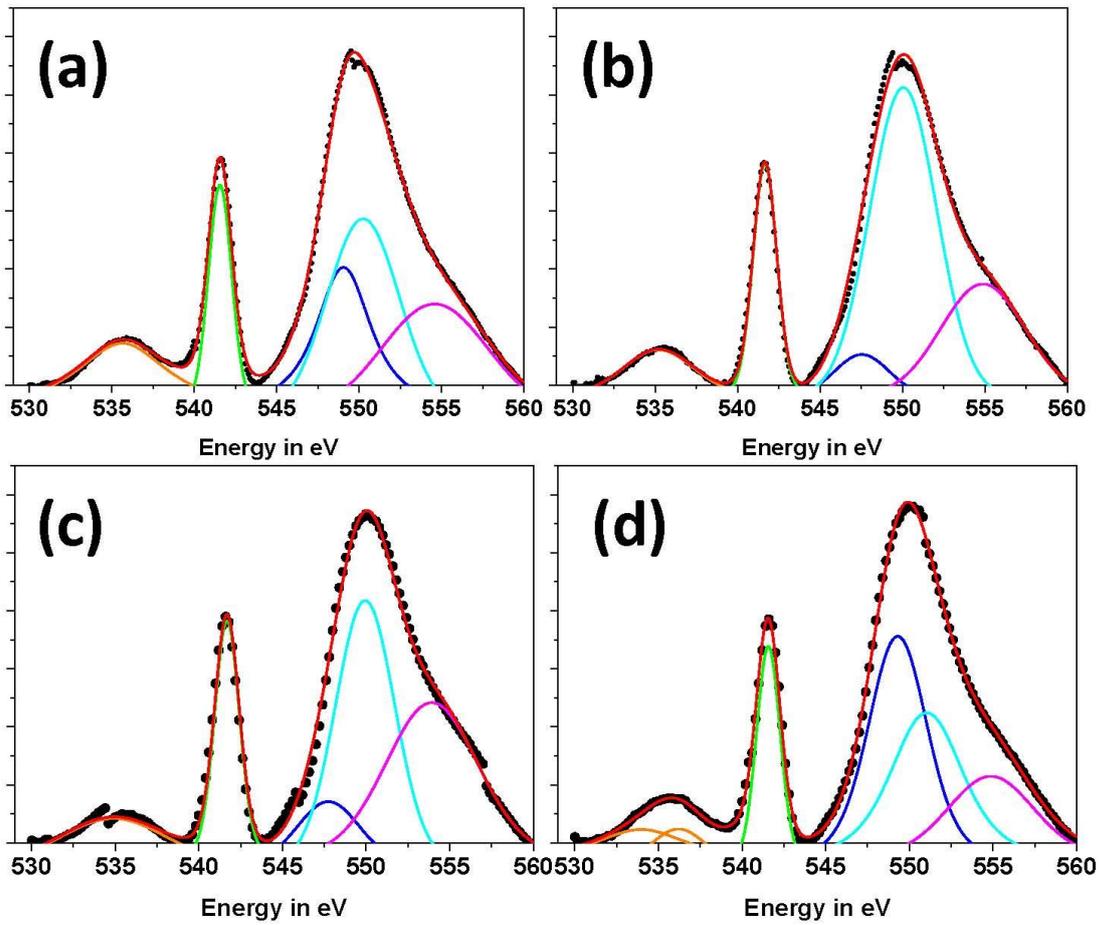

Figure 7: O K edge NEXAFS spectra (solid circles) with corresponding voigt fitted spectra (solid lines) of (a) DPPC; (b) Na:K-100:0, (c) Na:K-70:30 and (d) Na:K-50:50 samples

| Peak Number | Peak Colour | Dipalmitoylphosphatidylcholine Monolayer on Si(001) Peak Position (in nm) | | | | Tentative assignments |
|---|---|---|---|---|---|---|
| | | Pristine (No cations) | With Na:K=100:0 | With Na:K=70:30 | With Na:K=50:50 | |
| 1 | Red | 535.6 | 535.2 | 534.89 | 533.96 | C=O ($\pi^*$) |
| 2 | Red | | | | 536.2 | C-O ($\pi^*$)/C=O ($\pi^*$) |
| 3 | Green | 541.58 | 541.64 | 541.68 | 541.57 | C-O ($\sigma^*$)/Si-O |
| 4 | Blue | 549.05 | 547.53 | 547.75 | 546.58 | C=O ($\sigma^*$) |
| 5 | Cyan | 550.25 | 550.04 | 549.95 | 549.30 | P-O ($\sigma^*$) |
| 6 | Magenta | 554.58 | 554.78 | 553.93 | 554.86 | P=O ($\sigma^*$) |

Table 2: Voigt fitted peaks of NEXAFS spectra with peak assignments of the samples.

Table 2 shows that, like the changes in rigidity and angular stress, the variation of the P=O bond with ion ratio is also nonlinear. It has been observed earlier that at high pH of the aqueous sub-phase, the Na cations of the NaOH used to raise the pH can interact with more than one carboxyl headgroup of a lipid monolayer floating on the water surface [24]. This indicated an extended electrostatic field created by the cations to which the carboxyl anions respond. Such a field is inconsistent with the screened Coulomb potential of dispersed cations but can be explained by a dense layer of cations beneath the lipid monolayer, somewhat similar to the ionic lattices observed with some divalent cations [25].

The present results show further that, on varying the cation charge density linearly the phosphoryl anions respond through nonlinear variations in bonding, leading to the molecule-level and monolayer-level nonlinear variations with the cation ratio. This is again consistent with an extended electrostatic field of a dense, organized layer of cations. The observed variations point to the spatial variations in this field due to the localized decrease in field strength from the lower charge density of the $K^+$ ion. In absence of structural data in the monolayer plane, we are restrained from making more precise statements.

This variation in the field may explain the presence of the split in the C=O π-bond peak, signifying two different bonds with distinct energy values. At higher values of the field intensity the C=O bond may undergo more charge localization and polarization thereby reducing the bond energy while at lower fields the bond may retain its covalent nature and have a higher energy. Such a distinction is expected to be most pronounced at equal proportions of the two cations, as is observed in fact.

The P-O bond, carrying the charge of the phosphoryl group, presents another interesting aspect. The bond is seen to become progressively weak with the introduction of the cations and the drop becomes most noticeable as the K percentage increases from 30 to 50. According to Fajans rule [26] the $Na^+$-$O^-$ interaction has a polar bond-like nature due to the distortion of the electron clouds arising from the size mismatch of the partners, while this is reduced to a more electrovalent bonding as K is introduced and increased as the partners now match in size. The weakening of the associated P-$O^-$ bond is due to the localization of the negative charge caused by the cations in general and more by the dominantly electrovalent interaction of the $K^+$ ion in particular.

4. Conclusions

Two-dimensional rigidity of Langmuir monolayers of Dipalmitoylphosphatidylcholine (DPPC), a major phospholipid component of the cell membrane, was found to vary nonlinearly with a linear variation of the $[Na^+]/[K^+]$ ratio in the aqueous subphase, as found from Surface Pressure versus Specific Molecular Area Isotherm studies. These monolayers, were then transferred onto hydrophilized Si(001) substrates and studied through X-Ray Reflectivity (XRR) and Near edge x-ray absorption fine structure (NEXAFS) spectroscopy at the O K-edge. Results obtained from the former also showed nonlinear variation in the angular stress between the headgroups and tails within the lipid molecules, whereas the latter showed nonlinear variations in the phosphoryl bonds, with linear variation in the cation ratio. These results indicate an extended electrostatic field at the monolayer-subphase interface with spatial variations emerging as the cation ratio is changed. Such a scenario, in turn, is inconsistent with a dispersion of cations each isolated within its solvation sphere and suggests a dense and spatially structured layer of cations at the interface. The phosphoryl-cation bonding is also found to be affected by Fajans rule, which in turn reduces the P-O bond strength with introduction of ions in general and of K ions in particular. These results have important implications for all interfacial assemblies of charges at high concentrations. In particular, they can play a crucial role in understanding the interfacial dynamics of the cell membrane where large concentration ratios of Na and K exist on both sides of the membrane.

## 5. Acknowledgements:

BB gratefully acknowledge the support of Council of Scientific & Industrial Research (CSIR), India, for providing Senior Research Fellowship (File no: 09/028(1073)/2018-EMR-I ). The authors gratefully acknowledge Elettra Sincrotrone Trieste for beam time (proposal 20215097). We thank Swapnasopan Datta for important inputs, especially regarding the application of Fajans rules.

## 6. References:

[1] R. Mancinelli, A. Botti, F. Bruni, M. A. Ricci, and A. K. Soper, *Hydration of Sodium, Potassium, and Chloride Ions in Solution and the Concept of Structure Maker/Breaker*, J. Phys. Chem. B **111**, 13570 (2007).

[2] S. Vaitheeswaran and D. Thirumalai, *Hydrophobic and Ionic Interactions in Nanosized Water Droplets*, J. Am. Chem. Soc. **128**, 13490 (2006).

[3] E. Hawlicka and D. Swiatla-Wojcik, *MD Simulation Studies of Selective Solvation in Methanol−Water Mixtures: An Effect of the Charge Density of a Solute*, J. Phys. Chem. A **106**, 1336 (2002).

[4] S. T. Moin, T. S. Hofer, A. B. Pribil, B. R. Randolf, and B. M. Rode, *A Quantum Mechanical Charge Field Molecular Dynamics Study of $Fe^{2+}$ and $Fe^{3+}$ Ions in Aqueous Solutions*, Inorg. Chem. **49**, 5101 (2010).

[5] M. G. Giorgini, K. Futamatagawa, H. Torii, M. Musso, and S. Cerini, *Solvation Structure around the $Li^+$ Ion in Mixed Cyclic/Linear Carbonate Solutions Unveiled by the Raman Noncoincidence Effect*, J. Phys. Chem. Lett. **6**, 3296 (2015).

[6] I. Gladich, S. Chen, H. Yang, A. Boucly, B. Winter, J. A. Van Bokhoven, M. Ammann, and L. Artiglia, *Liquid–Gas Interface of Iron Aqueous Solutions and Fenton Reagents*, J. Phys. Chem. Lett. **13**, 2994 (2022).

[7] P. Dullinger and D. Horinek, *Solvation of Nanoions in Aqueous Solutions*, J. Am. Chem. Soc. (2023).

[8] B. Brönsted, F. Fajans, G. Hevesy, B. Bjerrum, A. W. Porter, F. I. G. Rawlins, and K. Fajans, *General Discussion*, Trans. Faraday Soc. **23**, 375 (1927).

[9] K. Fajans and O. Johnson, *Apparent Volumes of Individual Ions in Aqueous Solution[1]*, J. Am. Chem. Soc. **64**, 668 (1942).

[10] W. Kinzel and M. Schick, *Extent of Exponent Variation in a Hard-Square Lattice Gas*

*with Second-Neighbor Repulsion*, Phys. Rev. B **24**, 324 (1981).

[11] R. Subramaniam, S. Lynch, Y. Cen, and S. Balaz, *Polarity of Hydrated Phosphatidylcholine Headgroups*, Langmuir **35**, 8460 (2019).

[12] A. Raudino and D. Mauzerall, *Dielectric Properties of the Polar Head Group Region of Zwitterionic Lipid Bilayers*, Biophys. J. **50**, 441 (1986).

[13] S. Nannarone et al., *The BEAR Beamline at Elettra*, AIP Conf. Proc. **705**, 450 (2004).

[14] D. Wang, D. H. De Jong, A. Rühling, V. Lesch, K. Shimizu, S. Wulff, A. Heuer, F. Glorius, and H. J. Galla, *Imidazolium-Based Lipid Analogues and Their Interaction with Phosphatidylcholine Membranes*, Langmuir **32**, 12579 (2016).

[15] J. V. N. Ferreira, T. M. Capello, L. J. A. Siqueira, J. H. G. Lago, and L. Caseli, *Mechanism of Action of Thymol on Cell Membranes Investigated through Lipid Langmuir Monolayers at the Air–Water Interface and Molecular Simulation*, Langmuir **32**, 3234 (2016).

[16] S. Baoukina, L. Monticelli, S. J. Marrink, and D. P. Tieleman, *Pressure−Area Isotherm of a Lipid Monolayer from Molecular Dynamics Simulations*, Langmuir **23**, 12617 (2007).

[17] B. Bawali, J. Saha, M. K. Mukhopadhyay, and A. Datta, *Effect of Na/K Ratio on Fluidization of Lipid Monolayers*, Dae Solid State Phys. Symp. 2019 **2265**, (2020).

[18] L. G. Parratt, *Surface Studies of Solids by Total Reflection of X-Rays*, Phys. Rev. **95**, 359 (1954).

[19] D. F. Kienle, J. V. De Souza, E. B. Watkins, and T. L. Kuhl, *Thickness and Refractive Index of DPPC and DPPE Monolayers by Multiple-Beam Interferometry*, Anal. Bioanal. Chem. **406**, 4725 (2014).

[20] *NEXAFS Spectroscopy - Joachim Stöhr - Google Books*, https://books.google.co.in/books?hl=en&lr=&id=F5bsCAAAQBAJ&oi=fnd&pg=PA1&ots=6yIFVf0yga&sig=ilCsbhnL2aiVYofteMi0KHi1oA0&redir_esc=y#v=onepage&q&f=false.

[21] S. Mukherjee, A. Datta, A. Giglia, N. Mahne, and S. Nannarone, *Relating Structure with Morphology: A Comparative Study of Perfect Langmuir–Blodgett Multilayers*, Chem. Phys. Lett. **451**, 80 (2008).

[22] G. Küper, R. Chauvistré, J. Hormes, F. Frick, M. Jansen, B. Lüer, and E. Hartmann, *Phosphorus K Shell Photoabsorption Spectra of the Oxides $P_4O_6$, $P_4O_{10}$, $P(C_6H_5O)_3$ and $PO(C_6H_5O)_3$*, Chem. Phys. **165**, 405 (1992).

[23] M. Gliboff et al., *Orientation of Phenylphosphonic Acid Self-Assembled Monolayers on a Transparent Conductive Oxide: A Combined NEXAFS, PM-IRRAS, and DFT Study*, Langmuir **29**, 2166 (2013).

[24] A. Datta, J. Kmetko, A. G. Richter, C. J. Yu, P. Dutta, K. S. Chung, and J. M. Bai, *Effect of Headgroup Dissociation on the Structure of Langmuir Monolayers*, Langmuir **16**, 1239 (1999).

[25] J. Kmetko, A. Datta, G. Evmenenko, and P. Dutta, *The Effects of Divalent Ions on Langmuir Monolayer and Subphase Structure: A Grazing-Incidence Diffraction and*

*Bragg Rod Study*, J. Phys. Chem. B **105**, 10818 (2001).

[26] K. Fajans, *Struktur Und Deformation Der Elektronenhüllen in Ihrer Bedeutung Für Die Chemischen Und Optischen Eigenschaften Anorganischer Verbindungen*, Naturwissenschaften **11**, 165 (1923).